\documentclass[prd,showpacs,showkeys,preprint,12pt]{revtex4}
\usepackage{amsmath,amsthm,amsfonts,amssymb}
\usepackage[mathcal]{eucal}
\usepackage{mathrsfs}
\usepackage{graphicx}
\usepackage{dcolumn}
\usepackage{latexsym}
\usepackage{epsfig}
\usepackage{bm}
\usepackage[all]{xy}

\usepackage[english]{babel}
\usepackage{amsmath,amsthm,amsfonts,amssymb}
\usepackage{graphicx}
\usepackage{dsfont}

\DeclareMathOperator{\e}{e}
\DeclareMathOperator{\sech}{sech}
\DeclareMathOperator{\mx}{max}

\newcommand{\ie}{\begin{equation}}
\newcommand{\fe}{\end{equation}}

\newcommand\fverb{\setbox\fverbbox=\hbox\bgroup\verb}
\newcommand\fverbdo{\egroup\medskip\noindent%
            \fbox{\unhbox\fverbbox}\ }
\newcommand\fverbit{\egroup\item[\fbox{\unhbox\fverbbox}]}
\newbox\fverbbox

\def\text#1{\mbox{#1}}

\begin{document}
\title{Gauge fields in a string-cigar braneworld}
\author{F. W. V. Costa$^{a,b}$}

\author{J. E. G. Silva$^{a}$}

\author{D. F. S. Veras$^{a}$}

\author{C. A. S. Almeida$^{a}$}

\address{$^{a}$Universidade Federal do Cear\'a (UFC), Departamento de F\'{\i}sica, Campus do Pici, Caixa Postal 6030, 60455-760, Fortaleza, Cear\'{a}, Brazil}

\address{$^{b}$FAFIDAM, Universidade Estadual do Cear\'a \\ Limoeiro do Norte, Cear\'a, Brazil}



\begin{abstract}
In this work, we investigate the properties of the Abelian gauge vector field in the background of a string-cigar braneworld. Both the thin and thick brane limits are considered. The string-cigar scenario can be regarded as an interior and exterior string-like solution. The source undergoes a geometric Ricci flow which is related to a variation of the bulk cosmological constant. The Ricci flow changes the width and amplitude of the massless mode at the brane core and recovers the usual string-like behavior at large distances. By means of suitable numerical methods, we attain the Kaluza-Klein (KK) spectrum for the string-like and the string-cigar models. For the string-cigar model, the KK modes are smooth near the brane and their amplitude are enhanced by the brane core. Furthermore, the analogue Schr\"{o}dinger potencial is also regulated by the geometric flow.

\end{abstract}


\maketitle

\section{Introduction}

The braneworld models became an active research area in High Energy Physics in the last years. The seminal works of Randall-Sundrum (RS) \cite{Randall:1999ee,Randall:1999vf} introduced an infinite extra dimension by means of a warped compactification. Thereafter, several models explored both the field \cite{Kehagias:2000au,Bazeia:2004dh} as cosmological  properties \cite{Csaki:1999mp,Binetruy:1999hy} of the RS models. In six dimensions, by assuming a static and axial symmetry for the bulk, the geometry of the brane is similar to cosmic string space-time. Then, these models are called string-like braneworlds \cite{Olasagasti:2000gx, Cohen:1999ia, Gregory:1999gv, Oda2000a, Oda:2000zc, Liu:2007gk, Gherghetta:2000qi}.

Even though the string-like models have the advantage of traping free gauge fields \cite{Oda2000a,Oda:2000zc} and minimally coupled Dirac fermions \cite{Liu:2007gk}, they present some issues about the core of the source. The Gherghetta-Shaposhnikov (GS) model \cite{Gherghetta:2000qi}, for instance, does not satisfy the metric regularity conditions at the brane nor the dominant energy condition \cite{Tinyakov:2001jt}. This is due to the metric proposed be only a vacuum solution of the Einstein equations. The GS model can be regarded as an infinitely thin string-like model.

In order to study the source properties and to suppress the anomalies of the string-like models, Giovannini \textit{et al} proposed a braneworld scenario constructed from an Abelian vortex \cite{Giovannini:2001hh}. The solution found numerically satisfies all the regularity and energy conditions, but the analytical solution is still lacking \cite{Giovannini:2001hh}. Moreover, the same solutions leading to gravity localization, also lead to the localization of the gauge zero mode \cite{Giovannini_Gauge-6D-PRD, Giovannini_Gauge-6D-CQG}. Afterward, de Carlos-Moreno proposed a supersymmetric model without bulk cosmological constant free of regularity problems \cite{deCarlos:2003nq}, whereas Papantonopoulos \textit{et al} regularized the conical behavior near the brane by adding a ring-like structure at the brane \cite{Papantonopoulos:2007fk}. More recently, Silva-Almeida proposed a resolution scheme based in an effective conifold transition in the internal space \cite{Silva:2011yk}. The resolution of the conical singularity at the core of the brane provides a geometrical flow which smoothes the Kaluza-Klein (KK) modes for the scalar, gauge and Dirac fields \cite{Silva:2011yk,Costa:2013eua,Dantas:2013iha}.

An analytical smooth string-like model was proposed and its gravitational KK modes were studied \cite{Silva:2012yj}. This thick solution extends the GS model being an interior and exterior string-like model. The near brane corrections to the geometry makes the model satisfy all the regularity and energy conditions \cite{Silva:2012yj}. Since this string-like model is built from a warped product between a $3-$brane and a particular steady solution of the Ricci flow, called Hamilton cigar soliton, the scenario is called the string-cigar model \cite{Silva:2012yj}. The evolution parameter of the Ricci flow yields to a varying bulk cosmological constant and changes the ratio among the components of the stress-energy tensor \cite{Silva:2012yj}. Moreover, the near brane correction provides a potential well around the brane for the gravitational KK modes \cite{Silva:2012yj}.

Once defined and studied this smoothed string-like braneworld, it is worth to analyse the behavior of the Standard Model fields on this scenario. In this article, we  study the features of the gauge vector field minimally coupled to this geometry. We show that the radial component has a richer dependence compared with the thin string-like solution presented on Ref. \cite{Oda:2000zc}. Moreover, the KK tower has a trapped massless $s-$wave state shifted from the origin due to the core behavior. The displacement of the massless mode happens because the brane-core is shifted from the origin, as in the Abelian vortex solution for higher winding numbers \cite{Giovannini:2001hh}. Nevertheless, the source and the massless mode approach to the origin as the geometric evolution parameter increases. The limit of large values of this parameter matches with the thin string results.

The dynamics for the massive modes, presented in a Sturm-Liouville equation \cite{Silva:2012yj}, is quite complex to be studied analytically. Then, we accomplish a numerical analysis to find the KK spectrum and the respective eingenfunctions. It turns out that exists a mass gap between the massless and the massive modes for both the thin string-like and string-cigar models. Moreover, the known linearly increasing KK masses are obtained.
The KK eingenfunctions are all smooth near the brane and they recover the usual string-like pattern asymptotically \cite{Oda:2000zc}. The string-cigar core enhances the amplitude of the KK modes near the brane. Performing the analogue Schr\"{o}dinger approach, it turns out that the geometric flow, provided by a geometric parameter, controls the high of the barrier and the width of the potential well.

This article is organized as follows: In the Sec. \ref{Sec-TheStringCigarScenario}, we review the properties of the thin string-like and string-cigar models as well as the changes provided by the Ricci flow on the brane core. In the Sec. \ref{gaugelocalization}, we investigate the gauge vector massless field and the KK spectrum and study the influence of the geometric flow in this scenario for the $s-$wave state. Further in this section, we analyse the analogue Schr\"{o}dinger potential behavior upon this flow. Moreover, some conclusions and perspectives are outlined in Sec. \ref{Sec-Conclusion}.


\section{The String-Cigar Scenario}
\label{Sec-TheStringCigarScenario}

In this section, we briefly review the construction of the string-cigar model \cite{Silva:2012yj}. A complete description of the model, as well further details are presented in Ref. \cite{Silva:2012yj}.
Firstly, let us define a string-like spacetime. Let $\mathcal{M}_{6}$ a spacetime that can be split as $\mathcal{M}_{6}=\mathcal{M}_{4}\times \mathcal{M}_{2}$, where $\mathcal{M}_{4}$ is a $3-$brane embedded in $\mathcal{M}_{6}$ and $\mathcal{M}_{2}$ is a two dimensional Riemannian space. A string-like static braneworld  is a $\mathcal{M}_{6}$ with axial symmetry.  A warped metric ansatz for this axisymmetric spacetime is \cite{Olasagasti:2000gx, Cohen:1999ia, Gregory:1999gv, Oda2000a, Oda:2000zc, Gherghetta:2000qi}
\begin{eqnarray}
\label{metricansatz}
ds^{2}_{6} & =  & \sigma(r)\eta_{\mu\nu}dx^{\mu}dx^{\nu}+dr^{2} + \gamma(r) d\theta^{2},
\end{eqnarray}
where $\eta_{\mu\nu}$ is the induced metric brane and $\sigma$ and $\gamma$ are the so-called warp factors. In order to the brane possess a regular geometry, we assume that the warp functions satisfy the regularity conditions, namely \cite{Gherghetta:2000qi,Giovannini:2001hh}
\begin{eqnarray}
 \sigma(0)=1 &, & \sigma'(0)= 0\\
 \gamma(0)=0 & , &(\sqrt{\gamma})'(0)=1,
\end{eqnarray}
where the prime $(')$ denotes the derivative according to $r$ variable.

An example of a string-like geometry is the GS model \cite{Gherghetta:2000qi}, where $\mathcal{M}_{2}=\mathcal{S}^{2}$ is the two-dimensional disk of radius $R_{0}$ and the metric is given by the components \cite{Gherghetta:2000qi}
\begin{equation}
\label{thinstringbrane}
\sigma(r) = \e^{-c r} \hspace{0.5cm} \text{and} \hspace{0.5cm} \gamma(r)=R_{0}^{2}\sigma(r),
\end{equation}
where $c\in \mathbb{R}$ is a constant. The constant $c$ is related to the bulk cosmological constant $\Lambda$ by \cite{Gherghetta:2000qi}
 \begin{equation}
\label{clambdarelation}
 c^{2}=-\frac{16\pi}{5M^{4}_{6}}\Lambda \hspace{0.3cm} \Rightarrow \hspace{0.3cm} \Lambda<0.
\end{equation}
Therefore, the GS model is a $AdS_{6}$ space time which geometry is an exterior solution for a thin string-like brane \cite{Gherghetta:2000qi}.

In the GS model, the relation between the Bulk Planck mass $M_{6}$ and the brane Planck mass $M_{4}$ is given by \cite{Gherghetta:2000qi}
\begin{equation}
\label{GSmassessrelation}
M_{4}^{2}=\frac{5\pi}{3}\frac{\mu_{\theta}}{-\Lambda}M_{6}^{4}.
\end{equation}
Then, in order to solve the hierarchy problem, GS model imposes a fine tuning between the bulk cosmological constant and the angular tension $\mu_{\theta}$, what yields to \cite{Gherghetta:2000qi}
\begin{equation}
\label{cosmologicalconstantconstrain}
-\Lambda \ll \mu_{\theta}.
\end{equation}
The inequality (\ref{cosmologicalconstantconstrain}) constrains the possible values for the constant $c$. In the RS model the curvature is set to be small
in order to guarantee the model be obtained from the Horava-Witten model \cite{Randall:1999ee}. For the GS model, $c$ can assume any value satisfying the conditions (\ref{clambdarelation}) and (\ref{cosmologicalconstantconstrain}).

Despite all these important features, the GS model does not satisfy all the regularity conditions at the origin \cite{Tinyakov:2001jt}. For $r=0$, instead of a 3-brane we obtain a 4-brane \cite{Gherghetta:2000qi}. Further, although the thin brane be flat, the curvature at the origin is non zero \cite{Gherghetta:2000qi}. Besides these geometrical issues, the source also does not satisfy all the energy conditions \cite{Tinyakov:2001jt}. The drawbacks of the GS model arise because it is only a vacuum exterior solution of the Einstein equations \cite{Gherghetta:2000qi}.

In order to solve these problems concerning the thin string-like branes, an interior and exterior solution for a string-like braneworld was proposed taking the cigar soliton as the transverse space $\mathcal{M}_{2}$ \cite{Silva:2012yj}. The cigar soliton is a solution of the geometrical Ricci flow whose equation is given by \cite{Silva:2012yj,hamilton1,Oliynyk:2005ak,Oliynyk:2007bv,Headrick:2006ti,Lashkari:2010iy,chow}
\begin{equation}
\label{ricciflow}
 \frac{\partial g_{ab}(\lambda)}{\partial \lambda}= -2 R_{ab}(\lambda).
\end{equation}
The Ricci flow (\ref{ricciflow}) defines a family of geometries evolving under a parameter $\lambda$. The metric for the cigar soliton can be written as \cite{Silva:2012yj,chow}
\begin{equation}
\label{cigarmetric2}
ds^{2}_{\lambda}=dr^{2}+\frac{1}{\lambda^{2}}\tanh^{2}{\lambda r} \hspace{0.15cm} d\theta^{2}.
\end{equation}
The evolution parameter $\lambda$ may be identified with the warp constant $c$  \cite{Silva:2012yj}, so that, choosing the warp metric components
\begin{equation}
\label{warpfunction}
\sigma(r,c) = \e^{-\left(cr - \tanh{cr}\right)}
\end{equation}
and
\begin{equation}
\label{angularmetric}
\gamma(r,c)=\frac{1}{c^{2}}\tanh^2(cr) \hspace{0.1cm} \sigma(r,c),
\end{equation}
it defines an axisymmetric braneworld called string-cigar model \cite{Silva:2012yj}.

The string-cigar model converges to the GS model far from the origin \cite{Silva:2012yj}.
Near the origin, the string-cigar geometry presents a conical behavior and it smoothes out the warp factor. As a result, all the regularity conditions are satisfied and thereby, the string-cigar geometry is a smooth interior and exterior string-like solution \cite{Silva:2012yj}. The string-cigar model not only smoothes the GS model near the brane but also provides a geometrical flow due to the variation of the bulk cosmological constant. Indeed, the components of the stress-energy tensor evolves under the geometrical flow. For small $c$, the source satisfies the dominant energy condition, whereas for great values $c$, only the weak energy condition is satisfied \cite{Silva:2012yj}. Moreover, the width and the position of the brane core also change with the flow. The source approaches to the origin and becomes narrower as $c$ increases. Therefore, high cosmological constant values situate the source around the origin what agrees with the GS model \cite{Gherghetta:2000qi}.

Furthermore, the string-cigar model also provides a solution for the hierarchy problem. Indeed, the relation between the masses are now given by \cite{Silva:2012yj}
\begin{eqnarray}
M^{2}_{4} & = & 2\pi M_{6}^{4}\int_{0}^{\infty}{\sigma(r,c)\sqrt{\gamma(r,c)}dr}\nonumber\\
          & = & \frac{2\pi M_{6}^{4}}{c}\int_{0}^{\infty}{\e^{-\frac{3}{2}(cr-\tanh{cr})}\tanh{cr} \: dr}\nonumber\\
          & \approx & 2\pi M_{6}^{4}\frac{1}{c^{2}}.
\label{planckmass}
\end{eqnarray}
Therefore, as in the RS model, due to the dominant energy condition and the hierarchy solution, we concern ourselves to the small $c$ analysis.

It is worthwhile to mention that the near brane corrrections shifts the maximum of the energy density from the origin \cite{Silva:2012yj}. Similar results were found by numerical analysis performed by Giovannini \textit{et al} for higher winding number Abelian vortex \cite{Giovannini:2001hh}. Besides, Kehagias argued that this displacement due to the conical behavior could be used to explain the cosmological problem \cite{Kehagias:2004fb}.


\section{Gauge Field Localization}
\label{gaugelocalization}

Once we presented how the Ricci flow changes the properties of the brane core, we proceed to analyse the effects of this geometrical flow upon the Abelian vector field. We begin with a $U(1)$ invariant vector field action, namely
\begin{equation}
\label{gaugefieldaction}
S =\int{d^{6}x\sqrt{-g} \hspace{0.075cm} g^{MN}g^{RS} F_{MR}F_{NS}},
\end{equation}
where $F_{MN}=\nabla_{M}A_{N}-\nabla_{M}A_{N}$. From the action (\ref{gaugefieldaction}), the equation of motion is obtained in a  straightforward way as
\begin{eqnarray}
\label{gaugefieldmotionequation}
 \frac{1}{\sqrt{-g}} \partial_{R}(\sqrt{-g} \hspace{0.075cm} g^{RM}g^{LN}F_{MN})= 0.
\end{eqnarray}
Let us consider the brane Lorentz gauge
\begin{equation}
\partial_{\mu}A^{\mu}=0
\end{equation}
and a purely radial field configuration, e.g., $A_{\theta}=0$, as usual \cite{Oda2000a,Oda:2000zc,Costa:2013eua}. In addition, since the $3-$brane is flat and
has an axial symmetry, the radial vector component $A_{r}$ should not depend on the brane coordinates, i.e., \cite{Costa:2013eua}
\begin{eqnarray}
\label{symmetrygaugecondition}
A_{r}=A_{r}(r,\theta)   &   \Rightarrow &   \partial_{\lambda}A_{r}(x^{\mu})=0.
\end{eqnarray}

Using the string-cigar metric and the gauge choice, the equation of motion (\ref{gaugefieldmotionequation}) takes the form
\begin{equation}
\label{motionequation1}
\left(\eta^{\mu\nu}\partial_{\mu}\partial_{\nu}+\frac{\sigma(r,c)}{\gamma(r,c)}{\partial_{\theta}}^{2}\right)A_{r}=0,
\end{equation}
\begin{equation}
\label{motionequation2}
\partial_{r}\left(\frac{\sigma^{2}(r,c)\sqrt{\gamma(r,c)}}{\gamma(r,c)}\partial_{\theta}A_{r}\right)=0
\end{equation}
and
\begin{equation}
\label{equationmotion3}
\Big(\eta^{\mu\nu}\partial{\mu}\partial{\nu}+\frac{\sigma}{\gamma}\partial_{\theta}^{2}+\frac{1}{\sqrt{\gamma}}\partial_{r}\sigma\sqrt{\gamma}\partial_{r}\Big) A_{\lambda}
=0.
\end{equation}

Performing the Kaluza-Klein decompositions \cite{Oda2000a}
\begin{equation}
\label{equationmotion4}
A_{\mu}(x^M)=\sum_{n, l=0}^{\infty}A_{\mu}^{(n, l)}(x^{\mu})\chi_n(r)Y_{l}(\theta)
\end{equation}
and
\begin{equation}
\label{equationmotion5}
A_{r}(x^M)=\sum_{l=0}^{\infty}A_{r}^{(l)}(x^{\mu})\xi(r)Y_{l}(\theta),
\end{equation}
the Eq. (\ref{motionequation1}) yields to
\begin{equation}
\label{motionequation6}
\left(\eta^{\mu\nu}\partial_{\mu}\partial_{\nu}- \frac{\sigma}{\gamma}l^2\right)A_{r}^{(l)}(x^{\mu})=0.
\end{equation}
Thus, in order to match Eq. (\ref{motionequation1}) with Eq. (\ref{symmetrygaugecondition}), we restrict ourselves to $s-$waves states, e.g., $l=0$ \cite{Costa:2013eua}. Further, the Eq. (\ref{motionequation2}) leads to a general solution to $\xi(r)$ as
\begin{equation}
\label{motionequation7}
\xi(r) = k{\gamma}^{1/2}\sigma^{-2},
\end{equation}
where $k$ is an simple integration constant \cite{Costa:2013eua}. It is worth to mention that the function $\xi(r)$ in the Eq. (\ref{motionequation7})  is a direct extension to that displayed by Oda \cite{Oda2000a}. Finally, using Eq. (\ref{motionequation7}), the Eq. (\ref{equationmotion3}) turns to be
 \begin{equation}
 \label{chiequation}
 \chi_n''(r)+\left(\frac{3}{2}\frac{\sigma'}{\sigma}+\frac{1}{2}\frac{{\beta}'}{\beta}\right)\chi_n'(r)+\frac{m_n^2}{\sigma}\chi_n(r)=0,
 \end{equation}
where $\beta(r,c) = \gamma(r,c)/\sigma(r,c)$. The Eq. (\ref{chiequation}) governs the behavior of the gauge field through the bulk. It is worth to mention that this equation is rather similar to the graviton radial equation, regardless the change of the factor $\frac{3}{2}$ by $\frac{5}{2}$ \cite{Gherghetta:2000qi}.

We now impose the following boundary conditions \cite{Oda2000a, Gherghetta:2000qi, Giovannini:2001hh}
\begin{equation}
\label{boundarycondition}
\chi_n'(0)=\lim_{r\rightarrow\infty}\chi_n'(r)=0,
\end{equation}
which yields the orthogonality relation between $\chi_{i}$ and $\chi_{j}$ given by \cite{Gherghetta:2000qi}
\begin{equation}
\label{orthogonalityrelation}
\int_{0}^{\infty}\sigma(r,c)^{\frac{3}{2}}\sqrt{\beta(r,c)}\chi_{i}\chi_{j}dr=\delta_{ij}.
\end{equation}

Furthermore, it is possible to transform the KK  equation (\ref{chiequation}) into a Schr\"{o}dinger-like equation. Consider the change of independent variable \cite{Silva:2012yj}
\begin{equation}
\label{changevariable}
z=z(r)=\int^{r}\sigma^{-1/2}dr'
\end{equation}
and of dependent variable
\begin{equation}
\label{changeofdependentvariable}
\chi_n(z)=\Omega(z)\Psi_n(z),
\end{equation}
where $\Omega = C_1\sigma^{-1/2}\beta^{-1/4}$, with $C_1$ a constant. The changes of variables (\ref{changevariable}) and (\ref{changeofdependentvariable}) transform the Eq. (\ref{chiequation}) into a Schr\"{o}dinger one with  $\Psi_n(z)$ fulfilling
\begin{equation}
\label{schrodigner}
-\ddot{\Psi}_n(z)+U(z)\Psi_n(z)=m_n^2\Psi_n(z),
\end{equation}
where $U(z)$ is given by
\begin{equation}
U(z)= \frac{1}{4} \left[ 2\frac{\ddot{\sigma}}{\sigma} - \left(\frac{\dot{\sigma}}{\sigma}\right)^2 + \frac{\ddot{\beta}}{\beta} - \frac{3}{4}\left(\frac{\dot{\beta}}{\beta}\right)^2 + \frac{\dot{\sigma}}{\sigma} \frac{\dot{\beta}}{\beta} \right].
\end{equation}
The over-dots refers to derivatives with respect to the $z$ coordinate.


\subsection{Massless mode}
\label{Sec-Massless_mode}

For $m=0$, a solution for the Eq. (\ref{chiequation}) is the following linear combination:
\begin{equation}
\label{masslesssolution}
\chi_{0}(r)=C+\tilde{C}\int^{r}{\sigma^{-\frac{3}{2}}\beta^{-\frac{1}{2}}dr'},
\end{equation}
where $C$ and $\tilde{C}$ are integration constants. Since the second function does not satisfy the orthogonality condition (\ref{orthogonalityrelation}), we choose the constant function $\chi_{0}=C$ as the solution of Eq. (\ref{chiequation}).

For the thin string-like scenario \cite{Oda2000a}, the analogue Schr\"{o}dinger equation reads
\begin{equation}
\label{thinschrodinger}
-\ddot{\Psi}_n + \frac{2}{\left(z+\frac{2}{c}\right)^{2}}\Psi_n = m_n^{2}\Psi_n,
\end{equation}
where $z=\frac{2}{c}(\e^{\frac{c}{2}r}-1)$. Note that the origin in the $r$ coordinate is mapped into the origin in the $z$ coordinate.
The solution of Eq.(\ref{thinschrodinger}) for $m = 0$ is given by
\begin{equation}
\label{thinzeromodeschrodinger}
\Psi_{0}(z)=A_{1}\left(z+\frac{2}{c}\right)^{-1}+A_{2}\left(z+\frac{2}{c}\right)^{2}.
\end{equation}
By making $A_{2}=0$ in Eq. (\ref{thinzeromodeschrodinger}), we obtain a normalizable massless
\begin{equation}
\label{thinmasslessmode}
\Psi_{0}(r)=\sqrt{\frac{5c}{2R_{0}}}\e^{-\frac{c}{2}r}.
\end{equation}
This KK massless mode is trapped at the brane because of the exponential factor. Moreover, the amplitude is bigger for the gauge massless mode than for the gravitons \cite{Gherghetta:2000qi}.

From the general Schr\"{o}dinger equation (\ref{schrodigner}) for $m=0$, the solution satisfying the orthogonality condition (\ref{orthogonalityrelation}) is given by
\begin{equation}
\label{masslessmode}
\psi_{0}(r,c)=N\sigma(r,c)^{\frac{1}{2}}{\beta(r,c)}^{\frac{1}{4}},
\end{equation}
where
\begin{equation}
N^2=\frac{1}{\int_{0}^{\infty}\sigma(r,c)^{\frac{5}{2}}{\beta(r,c)}dr}
\end{equation}
is a normalization constant.

\begin{figure}[h]
 \centering
    \includegraphics[width=0.65\textwidth]{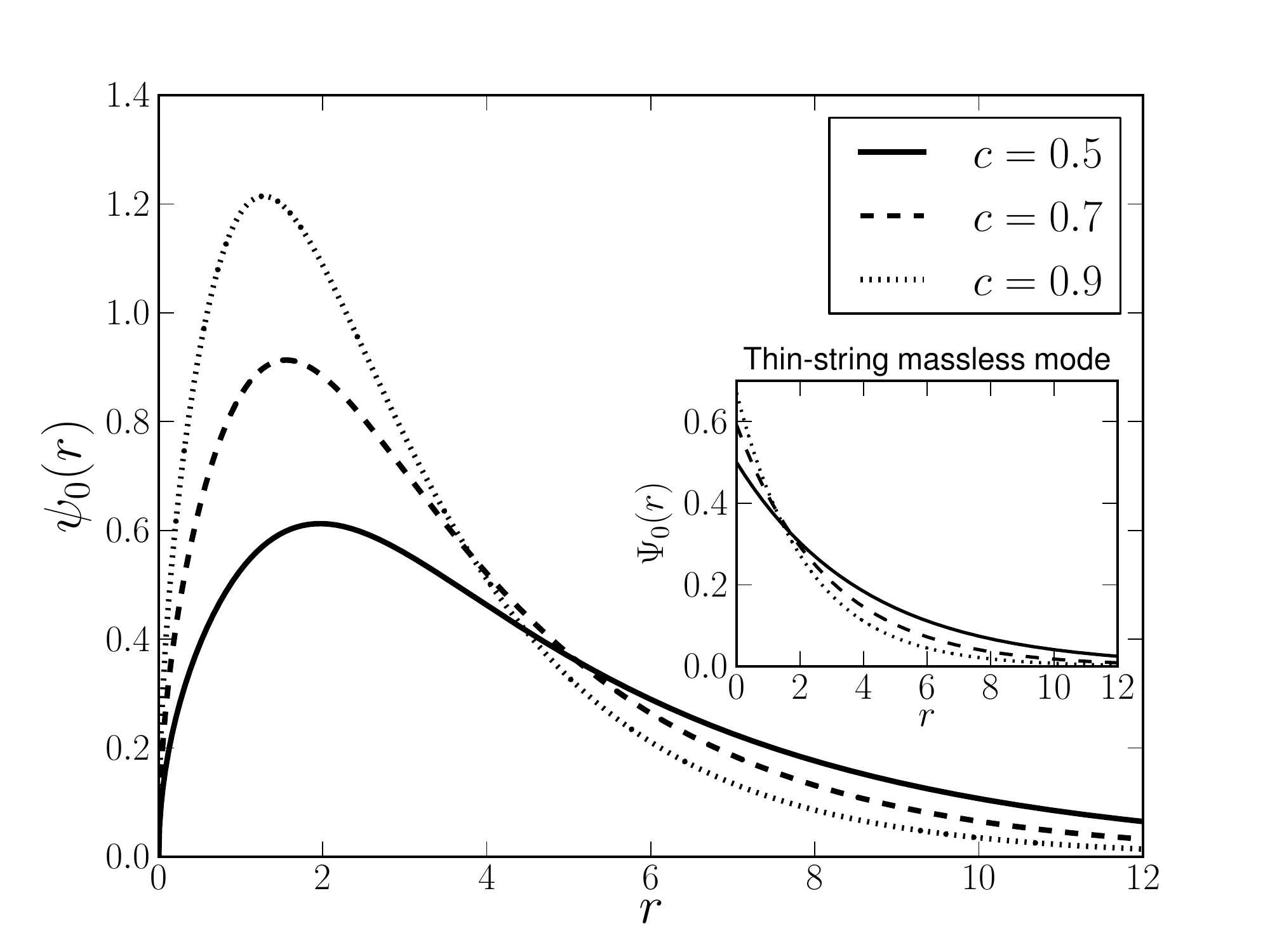}
 \caption{Massless mode for the string-cigar and thin string-like (on subgraph) scenarios.}
 \label{Fig-MasslessMode}
\end{figure}

We plot on figure (\ref{Fig-MasslessMode}) the massless mode for the string-cigar and for the  thin-string scenarios. These localized modes are responsible for the effective $3-$brane gauge field. The gauge massless modes are more concentrated at origin when compared with the graviton massless modes \cite{Gherghetta:2000qi, Silva:2012yj}. A worthwhile feature is the displacement of the massless mode from the origin in the string-cigar background. This behavior is also present in the energy density of the model (see Ref. \cite{Silva:2012yj}). Out of the core, the string-like exponential behavior dominates, whereas at the origin, the near core correction vanishes the mode. Matching these two regimes, there is a smooth peak which maximum is around the boundary of the brane core. Note that, for great values of the geometric parameter $c$ (and then, the bulk cosmological constant), the gauge massless mode in the string-cigar scenario tends to the thin string-like case.

\subsection{Massive modes}
Using the expressions for the metric factors (\ref{warpfunction}) and (\ref{angularmetric}), we obtain the KK equation from Eq. (\ref{chiequation}) in the form
\begin{equation}
\label{CompleteEquationChi}
\chi_n'' + c\left[-\frac{3}{2}\tanh^{2}cr+ \frac{\sech^{2}cr}{\tanh cr}\right]\chi_n' + \e^{(cr-\tanh{cr})}m_n^2\chi_n=0.
\end{equation}

Asymptotically, the Eq. (\ref{CompleteEquationChi}) recovers the thin-string model case, presented in Ref. \cite{Oda2000a}, as
\begin{equation}
\chi_n^{\prime \prime}(r) - \frac{3}{2}c\chi_n^{\prime}(r) + \e^{cr}m_n \chi_n(r) = 0,
\label{Oda-SturmLiouville_Eq}
\end{equation}
which general solution is \cite{Oda2000a}
\begin{equation}
\chi_n(r) = \frac{1}{N_n} \e^{\frac{3}{4} cr}\left[J_{3/2}\left(\frac{2m_n}{c} \e^{\frac{c}{2}r}\right) + \alpha_n Y_{3/2}\left(\frac{2m_n}{c} \e^{\frac{c}{2}r} \right) \right],
\label{Oda-Massive-Solution}
\end{equation}
where $N_n$ are normalization constants and $\alpha_n$ are constant coefficients determined by the boundary conditions. Looking at the gravitational case presented in Ref. \cite{Gherghetta:2000qi}, the graviton fluctuation $\phi$ has the radial solution
\begin{equation}
\phi_n(r) = \e^{\frac{5}{4} cr}\left[ B_1 J_{5/2}\left(\frac{2m_n}{c}\e^{\frac{1}{2}cr}\right) + B_2 Y_{5/2}\left(\frac{2m_n}{c}\e^{\frac{1}{2}cr}\right)\right],
\label{GS-Massive-Solution}
\end{equation}
where $B_1$ and $B_2$ are arbitrary constants. Thus, the massive modes of the gauge field have a higher amplitude near the brane but they spread less into the bulk when compared with the gravitons.

For the thin string-like model, there is a discontinuity between the massive and the massless mode. Indeed, by making the limit $m_{n}\rightarrow 0$ the
massive states converge to $\phi_{n}(r)=0$ not to $\phi_{0}(r)$. Further, transforming the Eq. (\ref{Oda-SturmLiouville_Eq}) into a Schr\"{o}dinger-like equation, we find that the KK Schr\"{o}dinger solutions has the form
\begin{equation}
\label{thinmassiveschrodinger}
\Psi_{m}(z)=\sqrt{\frac{2}{\pi}}\biggl[\frac{(A-mB\bar{z})\sin(m\bar{z})-(mA\bar{z} + B)\cos(m\bar{z})}{m \bar{z}}\biggr],
\end{equation}
where $A$ and $B$ are integration constants and $\bar{z} = z+\frac{2}{c}$. Since the KK solutions in Eq. (\ref{thinmassiveschrodinger}) are not defined for $m=0$, then we can not obtain the massless mode
continuously from the massive modes. The existence of this kind of mass gap was also found in negative tension braneworlds in five dimensions \cite{Giovannini-5D}.

Applying the boundary conditions (\ref{boundarycondition}) at some cut off point $r_{\mx}$, it is possible to obtain the KK spectrum.
In fact, the boundary conditions yield to the system of equations
\begin{eqnarray}
\label{boundaryconditionsystem}
 J_{\frac{1}{2}}\left(2\frac{m_{n}}{c}\right) + \alpha_n Y_{\frac{1}{2}}\left(2\frac{m_{n}}{c}\right)	&	=	&	0\\
  J_{\frac{1}{2}}\left(2\frac{m_{n}\e^{\frac{c}{2} r_{\mx}}}{c}\right) + \alpha_n Y_{\frac{1}{2}}\left(2\frac{m_{n}\e^{\frac{c}{2} r_{\mx}}}{c}\right)	&	=	&	0.
\end{eqnarray}
The system above is difficult to be treated analytically.
However, for the small mass regime, i.e. $m_n \ll c$, the divergence in the Bessel function of second kind yields to $\alpha_{n}=0$. Hence, the mass spectrum can be obtained from the equation \cite{Oda2000a}
\begin{equation}
J_{\frac{1}{2}}\left(2\frac{m_{n}\e^{\frac{c}{2}r_{\mx}}}{c}\right)=0.
\end{equation}
Then, from the zeroes of the Bessel function $J_{\frac{1}{2}}(x_{n})$, where $x_{n}=\frac{2m_n}{c}\e^{\frac{c}{2}r_{\mx}}$, we obtain the massive spectrum as \cite{watson}
\begin{equation}
 m_{n}=\frac{c}{2}n\pi \e^{-\frac{c}{2}r_{\mx}}.
 \label{Eq_Spectrum-Exatc}
\end{equation}
Note that the gauge masses grow linearly with the discrete Bessel function root index $n\in\mathbb{N}$, as in the factorizable Kaluza-Klein model \cite{Oda2000a}. Therefore, for $(m_{n} \ll c)$, the spectrum is discrete and there is an exponential suppressed mass gap between the massless mode and the first massive mode given by
\begin{equation}
\Delta=m_{1}-m_{0}=\frac{c}{2}\pi \e^{-\frac{c}{2}r_{\mx}}.
\end{equation}


Once the general KK equation ($\ref{CompleteEquationChi}$), as well the boundary condition system (\ref{boundaryconditionsystem}), are difficult to solve analytically, we look for approximate solutions by numerical methods. Using the matrix method \cite{MatrixMethod} with second order truncation error, we obtain the complete KK spectrum and eigenfunctions for the gauge field on thin-string and string-cigar scenarios.

In order to obtain the gauge field KK spectrum on the thin-string scenario, we approximated the derivatives by finite differences in the grid points $r_j = jh$ with constant step-size $h = 0.01$. The domain used was $r \in [0.0, \hspace{0.1cm}11.0]$. For the string-cigar scenario, the numerical integration of Eq. ($37$) was performed on the domain $r \in [0.01, 11.01]$ in order to avoid the singularity at $r = 0$.

\begin{figure}[h]
 \centering
    \includegraphics[width=0.65\textwidth]{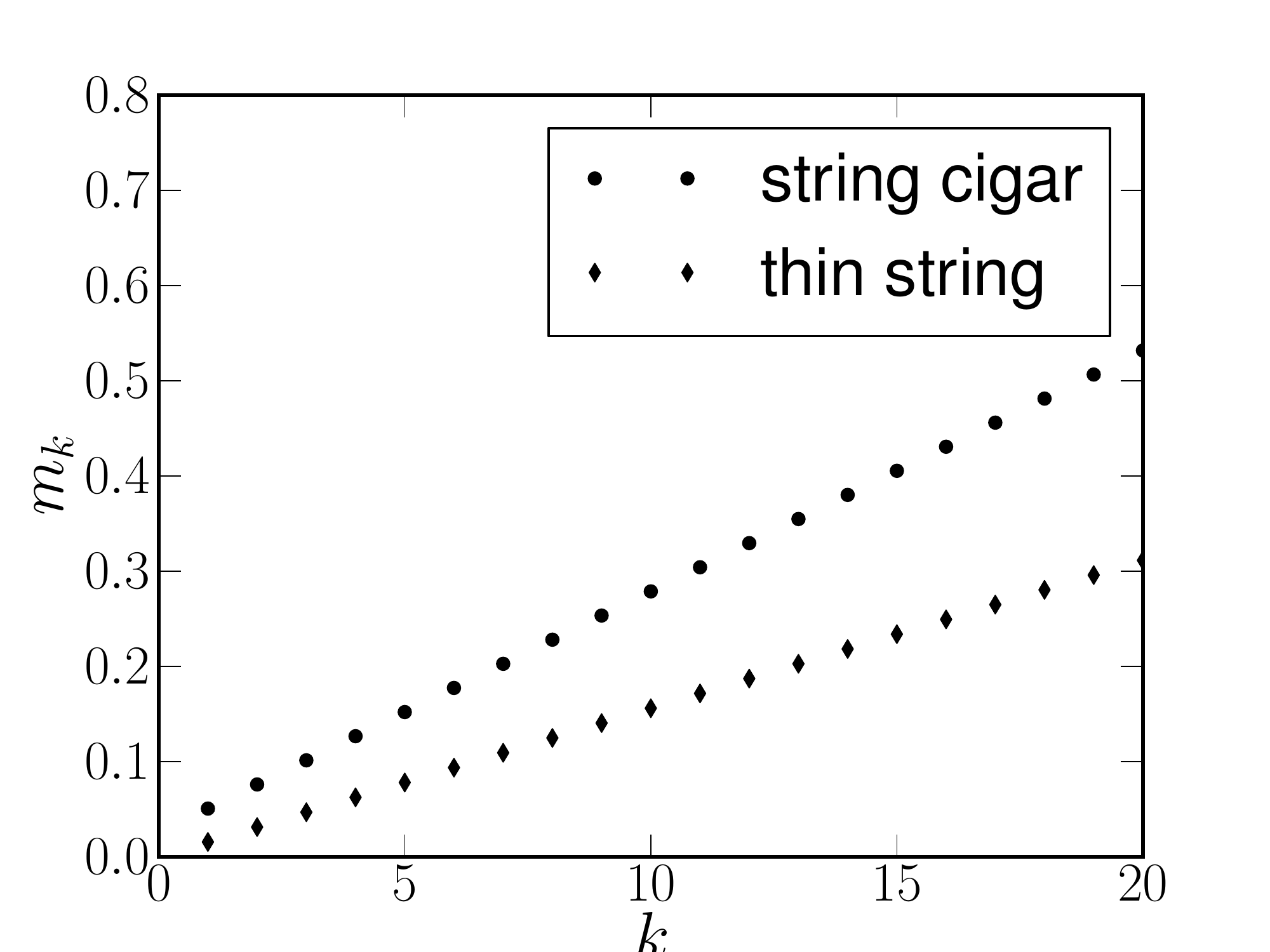}
 \caption{Kaluza-Klein spectrum of the gauge vector field in the thin-string and string-cigar braneworlds for $c = 0.8$.}
 \label{Fig-Spectra}
\end{figure}

We plot in the figure \ref{Fig-Spectra} the lowest mass eigenvalues $m_k$ ($k = 1,2,3\cdots$) for both braneworld models. Note that the usual linear behaviour from the Kaluza-Klein theories are reproduced, where the index $k$ is the Kaluza-Klein number.
Since the Eq. $(37)$ is rather complex to be solved analytically, an exact value for $m_k$ can not be obtained to compare with the numerical values. However, once the near-brane correction of the string-cigar braneworld enhances the amplitude of the eigenfunctions \cite{Diego}, it is expected the same effect for the spectrum.


The eigenfunctions were obtained for both models whose shape is plotted in the figure (\ref{Fig-Autofuncao}) for $c = 0.8$. Asymptotically, all the solutions behave as Eq. (\ref{Oda-Massive-Solution}) whereas near the origin the amplitude of the KK modes are greater in the smooth string-cigar scenario than in the thin string-like one. The string-cigar solutions behave as Bessel functions of first kind near the origin.

\begin{figure}
        \centering
                \includegraphics[width=0.5\textwidth]{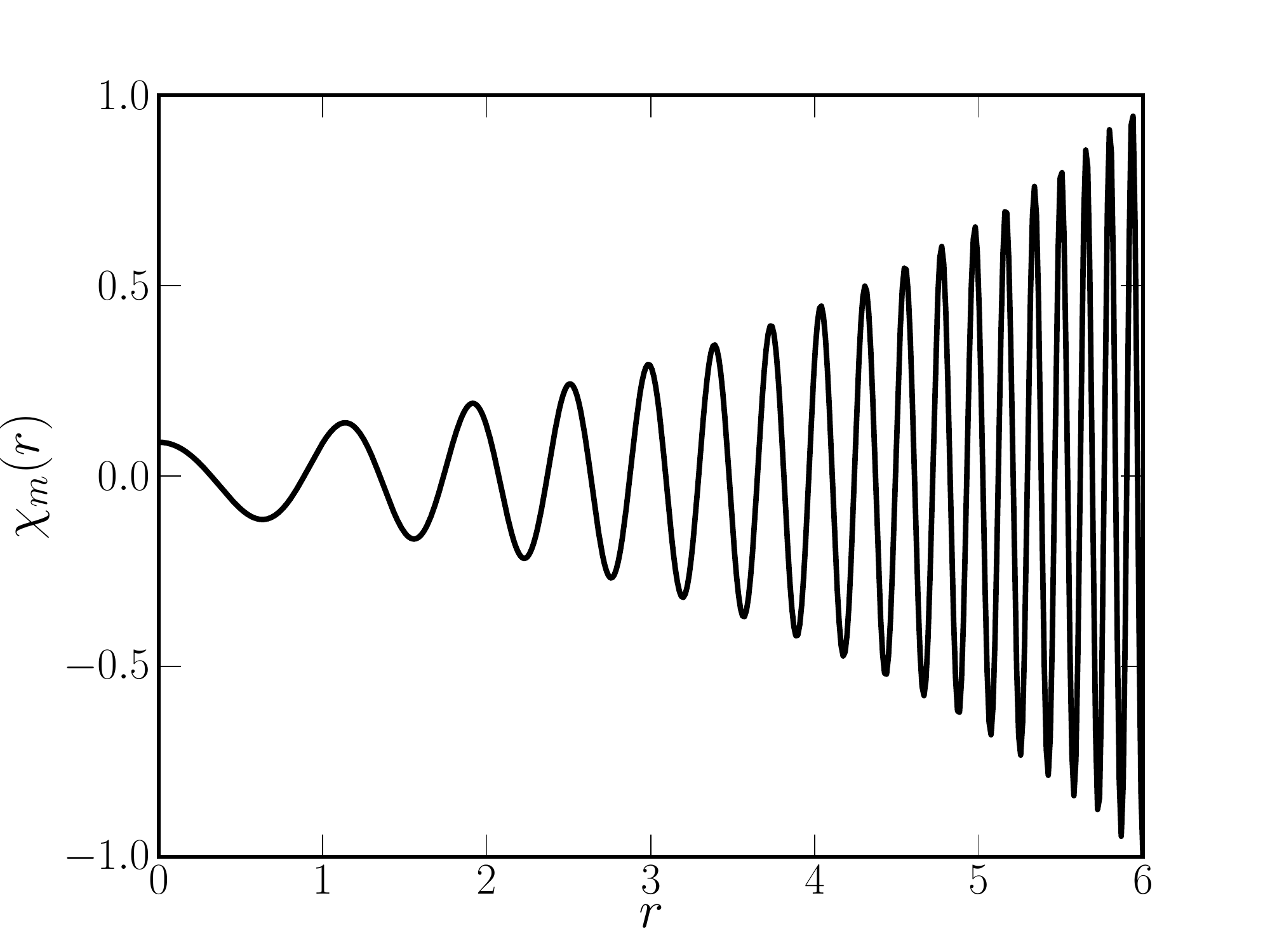}
                \caption{Gauge field KK numerical eigenfunction in the thin-string for $m = 0.259$ scenario.}
                \label{Fig-Autofuncao-Oda}
        ~ 
\end{figure}
\begin{figure}
                \includegraphics[width=0.5\textwidth]{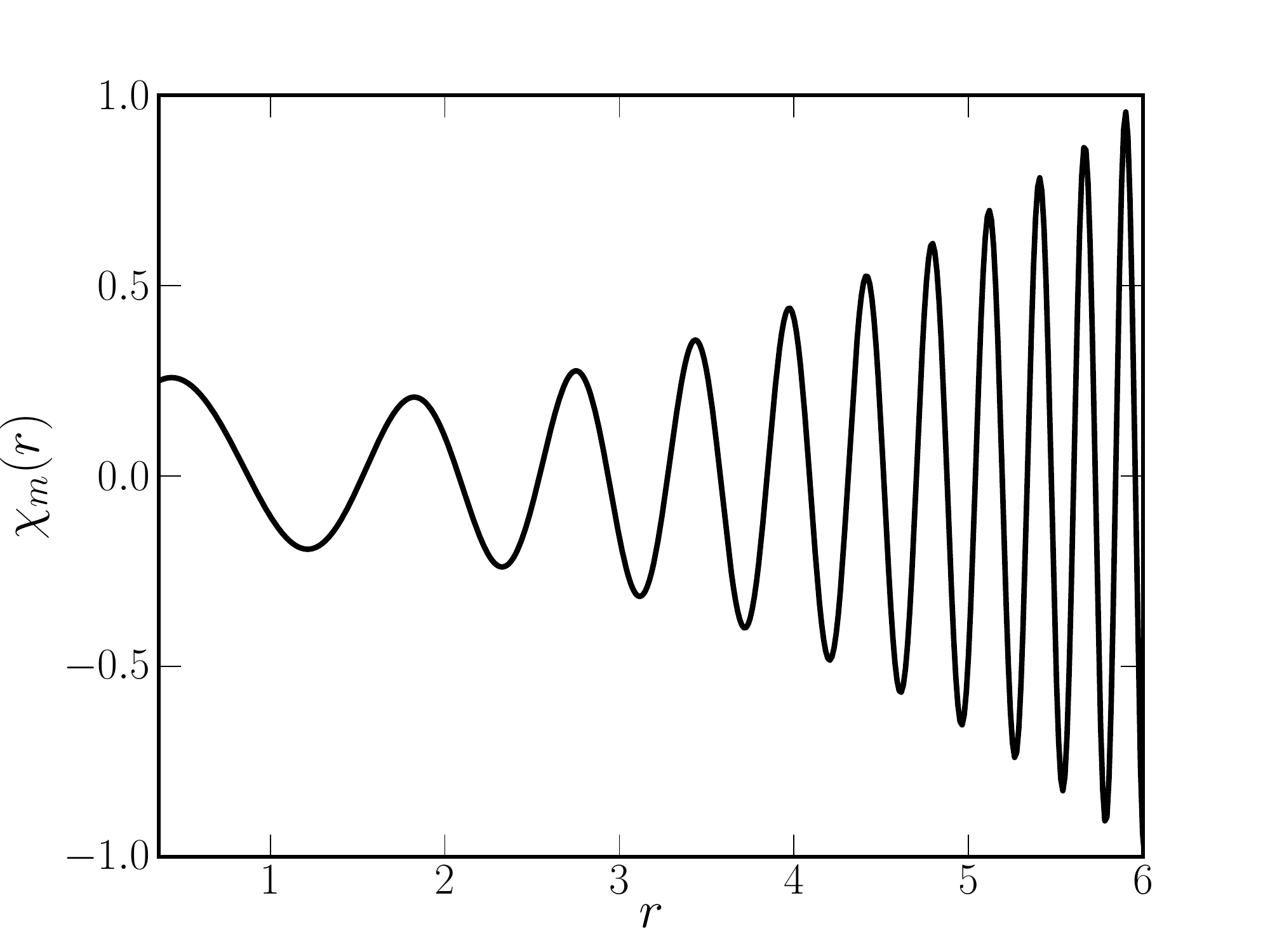}
                \caption{Gauge field KK numerical eigenfunction in the string-cigar for $m = 0.253$ scenario.}
                \label{Fig-Autofuncao-Charuto}
        ~ 

\end{figure}

\begin{figure}
        \centering
                \includegraphics[width=0.5\textwidth]{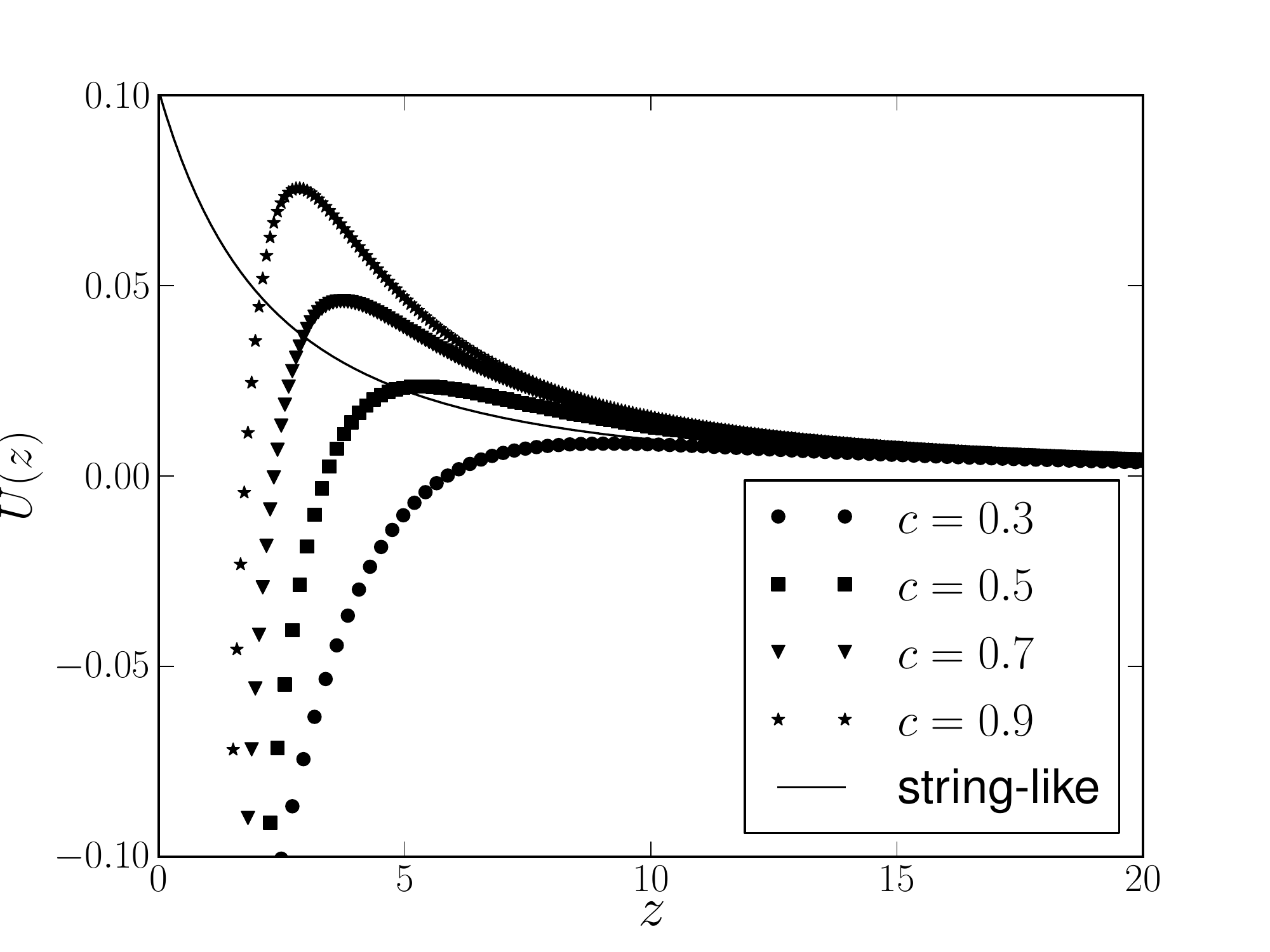}
                \caption{Numerical approximation of the quantum analogue potential $U(z)$ for some values of $c$. The thin line is a plot of the potential in the thin string-like background for $c = 0.45$. }
                \label{Fig-Potential}
        ~ 
\end{figure}
\begin{figure}
                \includegraphics[width=0.5\textwidth]{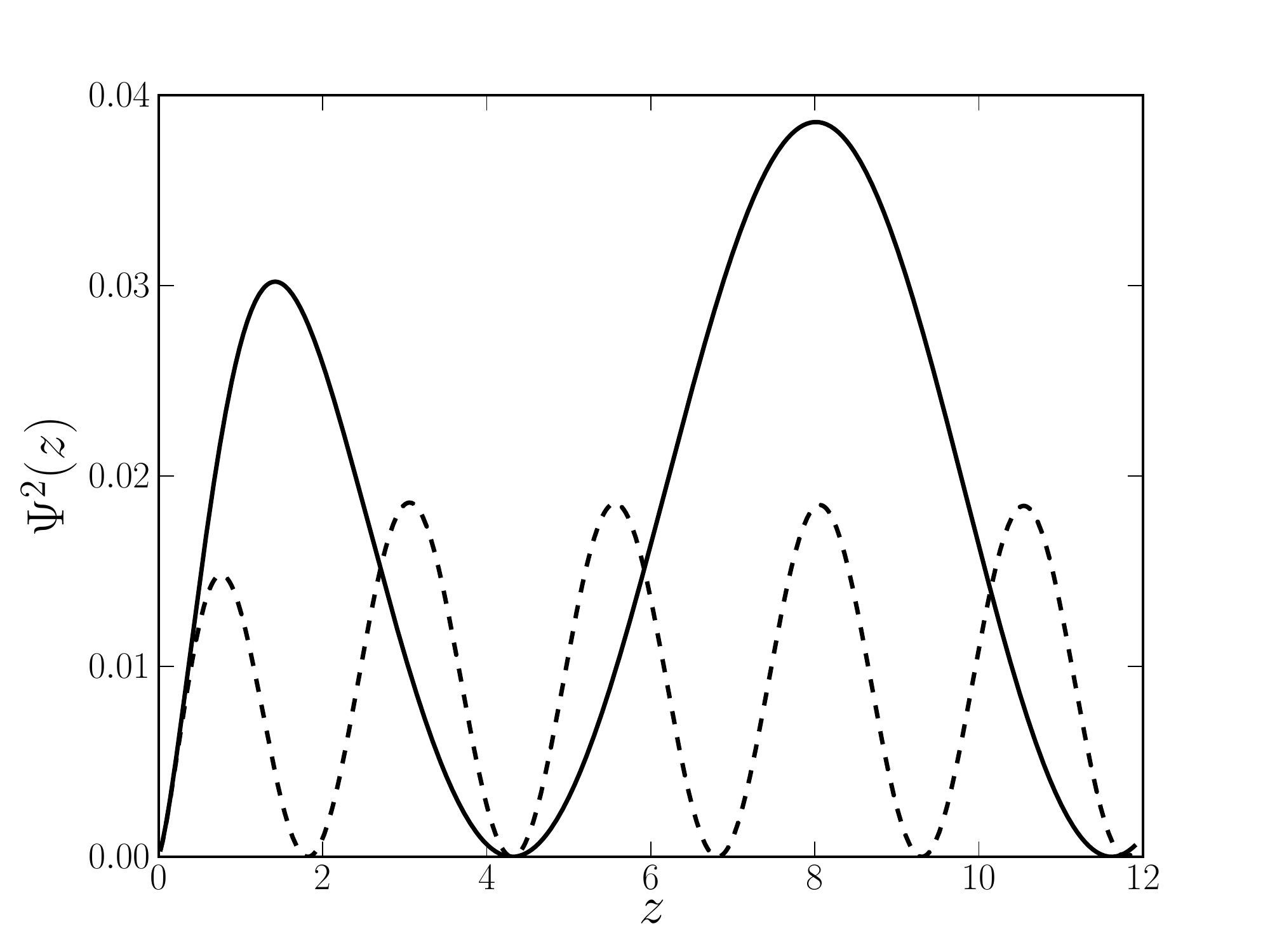}
                \caption{Numerical solutions of the Schr\"odinger-like equation for $c = 0.7$ and for small mass values $m = 0.45$ (dashed line) and $m = 1.27$ (thick line).}
                \label{Fig-FuncaoDeOnda}
        ~

\end{figure}

The massive states were also investigated numerically from the Schr\"odinger-like equation for the string-cigar scenario. The analogue quantum potential was constructed by numerical interpolation from the numerical integral (\ref{changevariable}). We have plotted the potential function $U(z)$ on the figure (\ref{Fig-Potential}) for some values of the evolution parameter. The potential well has a volcano-shape whose barrier increases and approaches to the origin with the increasing of $c$.
The Schr\"odinger-like equation (\ref{schrodigner}) was solved using the Numerov algorithm \cite{Numerov1}. We plotted in the figure (\ref{Fig-FuncaoDeOnda}) two wavefunctions for $c = 0.7$. The potential well influenced the first cycle of the wavefunctions and, for a moderate mass, the solutions rapidly oscillate. Modes with intermediate values of mass smoothly interpolate between the two solutions shown. Similar results were found for other values of $c$.




\section{Conclusions and Perspectives}
\label{Sec-Conclusion}

In this article we studied the features of the gauge vector field in the thin string-like and in the thick string-cigar model which evolves under a geometric Ricci flow.

The analysis was carried out for the $s-$wave states, i.e., $l=0$. The massless mode is localized and smoothed out compared to the massless mode in the thin string-like scenario. The maximum of this mode is displaced from the origin, likewise the stress energy components.
Asymptotically, the thin-string exponential behavior is recovered, whereas inside the core, the massless mode exhibits a conical behavior.

We have obtained the KK spectra for the string-like and string-cigar models by numerical techniques.
The well-known linear increasing behaviour was obtained, as well the massive gap between the massless mode and the first massive state in the thin-string background.
The massive eigenfunctions present a bigger amplitude near the origin in the string-cigar braneworld whereas asymptotically the KK modes exhibit the usual thin string-like behaviour. Thus, the string-cigar geometry provides a near brane correction to the KK modes.

We have also obtained numerically the analogue quantum potential. It possess a well known volcano-shape which width of the well and the high of the barrier are controlled by the geometric evolution parameter.

As perspectives we intent to obtain the corrections to the Coulomb potential \cite{Guo:2011qt}. Further, we propose to study the effects of the Ricci flow on the Dirac fermion field. Another interesting subject is the analysis of the massive KK modes for $l\neq 0$ configuration and the search of massive resonant modes.



\section{Acknowledgments}

The authors thank the Coordena\c{c}\~ao de Aperfei\c{c}oamento de Pessoal de N\'{\i}vel Superior (CAPES, grant 23038007567/2011-46), the Conselho Nacional de Desenvolvimento Cient\'{\i}fico   e Tecnol\'ogico (CNPq, grant 305766/2012-0), and Funda\c{c}\~ao Cearense de apoio ao Desenvolvimento Cient\'{\i}fico e Tecnol\'ogico (FUNCAP, grant BFP-026-00368.01.00/13) for financial support.‎ We would also like to express our thank to the anonymous referee for their enlightening questions and constructive comments.


\end{document}